\font\fourteenbf=cmbx12 at 14pt
\font\fourteenrm=cmr12 at 14pt
\font\fourteenit=cmti12 at 14pt
\font\twelvebf=cmbx12
\font\twelverm=cmr12
\font\twelveit=cmti12
\font\twelvesl=cmsl12
\font\tenbf=cmbx10
\font\tenrm=cmr10
\font\tenit=cmti10
\font\tenex=cmex10
\font\tensl=cmsl10
\font\ninebf=cmbx9
\font\ninerm=cmr9
\font\nineit=cmti9
\font\nineex=cmex9
\font\ninesl=cmsl9
\font\eightbf=cmbx8
\font\eightrm=cmr8
\font\eightit=cmti8
\font\eightex=cmex8
\font\eightsl=cmsl8
\font\sevenbf=cmbx7
\font\sevenrm=cmr7
\font\sevenit=cmti7
\font\sevenex=cmex7
\def\fourteen{\let\rm=\fourteenrm \let\bf=\fourteenbf \let\it=\fourteenit
            \let\sl=\fourteensl}
\def\twelve{\let\rm=\twelverm \let\bf=\twelvebf \let\it=\twelveit 
            \let\sl=\twelvesl}
\def\ten{\let\rm=\tenrm \let\bf=\tenbf \let\it=\tenit
         \let\ex=\tenex \let\sl=\tensl}
\def\nine{\let\rm=\ninerm \let\bf=\ninebf \let\it=\nineit
          \let\ex=\nineex \let\sl=\ninesl}
\def\eight{\let\rm=\eightrm \let\bf=\eightbf \let\it=\eightit
           \let\ex=\eightex \let\sl=\eightsl}
\def\seven{\let\rm=\sevenrm \let\bf=\sevenbf \let\it=\sevenit
           \let\ex=\sevenex } 
\hsize=6.0truein
\vsize=8.6truein
\baselineskip=15pt
\parindent=15pt
\twelve
\newif\ifdebug 
\newif\ifbacketedref 
\newif\ifpareqref 
\newif\ifpareqno 
\newif\ifeqnoperchapter 
\def\chapterfont{\twelvebf}

\def\titlefont{\twelvebf}
\def\centerline#1{\line{\hss{#1}\hss}}%
\def\title#1{{\titlefont\centerline{#1}\vskip 5mm}}
\def\authors#1{\centerline{#1} }
\def\DURHAM{{\parskip 2mm\tenit\baselineskip=13pt
    \centerline{Department of Mathematical Sciences}
    \centerline{University of Durham, Durham DH1 3LE, England}
}}

\def\Email#1{{\parskip 2mm\tenit\baselineskip=13pt\centerline{E-Mail: #1}}}

\def\abstract#1{{\tenrm\baselineskip=13pt
\parindent=0pt\vskip 1cm
\centerline{ABSTRACT}
\vbox{}
\vbox{\leftskip=12mm\rightskip=12mm#1}
}}
\newtoks\date 
\def\monthname{\relax\ifcase\month 0/\or January\or February\or
    March\or April\or May\or June\or July\or August\or September\or
    October\or November\or December\else\number\month/\fi}
\date={\monthname\ \number\day, \number\year}
\def\pubnum#1{{\twelverm\obeylines\everypar{\hfill} 
\rm DTP-#1

\the\date}
\vskip 5mm}
\newcount\chapternumber \chapternumber=0 
\newcount\sectionnumber \sectionnumber=0
\def\chapterreset{%
   \global\advance\chapternumber by 1%
   \ifeqnoperchapter \global\equanumber=0 \fi
   \sectionnumber=0 
}
\def\chapter#1{
{ \parindent=0pt \vglue 0.5cm
  \baselineskip=13pt
  \chapterreset
  {\chapterfont \number\chapternumber .\ #1 \hfill}
  \vglue 0.2cm
}}

\newcount\equanumber         \equanumber=0
\def\eqname#1{
  \def\Ename{\string#1}
  \relax 
  \ifnum\the\equanumber<0
    \xdef\LEQno{{\tenrm(\number-\equanumber)}}\global\advance\equanumber by -1
  \else \global\advance\equanumber by 1
    \ifdebug
       \xdef\LEQno{\Ename} 
    \else
      \ifeqnoperchapter
        \xdef\LEQno{\number\chapternumber .\number\equanumber} 
      \else
        \xdef\LEQno{\number\equanumber} 
      \fi
    \fi
  \fi
  \ifpareqref
    \xdef#1{{\rm(\LEQno)}\ }
  \else
    \xdef#1{{\rm Eq.\LEQno}\ }
 \fi
  \ifpareqno {\tenrm(\LEQno)} \else {\tenrm\LEQno} \fi 
}

\def\eqn#1{\eqno\eqname{#1}}

\newif\ifbacketedref
\newif\ifreferenceopen       \newwrite\referencewrite
\newdimen\referenceminspace  \referenceminspace=25pc
\newdimen\refindent          \refindent=30pt
\newcount\referencecount     \referencecount=0
\def\reffile{rf}
\immediate\openout\referencewrite=\reffile
\def\refout{\par \penalty-400 \vskip\chapterskip 
 \immediate\closeout\referencewrite
   \referenceopenfalse
   \vglue 0.5cm
   \baselineskip=13pt
   {\chapterfont References\hfil}
   \vglue 0.2cm
   \baselineskip=15pt
   \input \reffile
   }
\def\Textindent#1{\noindent\llap{#1\enspace}\ignorespaces}
\def\refitem#1#2{\par \hangafter=0 \hangindent=\refindent \Textindent{#1} #2}
\def\refnum#1{\global\advance\referencecount by 1 \def\Rname{\string#1}
\ifdebug\xdef#1{\Rname}\else\xdef#1{\the\referencecount}\fi
}
\def\refmark#1{\hbox{\raise1ex\hbox{{\eightrm%
   \ifbacketedref[#1]\else#1\fi}}}}

\def\Ref#1#2{\refnum#1\refmark{#1}%
  \immediate\write\referencewrite{\noexpand\refitem{#1.}{#2}}%
}
\def\REF#1#2{\refnum#1%
 \immediate\write\referencewrite{%
 \noexpand\refitem{#1.}{#2}}%
}

\newskip\headskip 	\headskip=8pt plus 3pt minus 3pt
\newskip\chapterskip    \chapterskip=\bigskipamount
\newskip\sectionskip     \sectionskip=\medskipamount
\def\ack{\par\penalty-100\medskip 
    \line{\twelverm\hfil ACKNOWLEDGEMENTS\hfil}\nobreak\vskip\headskip }
\newcount\Tableno     \Tableno=0
\def\figureinc{%
   \global\advance\figureno by 1%
}
\def\Table#1#2#3{\vskip 10mm%
{#3}
\global\advance\Tableno by 1%
\xdef#1{{\rm\number\Tableno}\ }%
\vskip 2mm
\centerline{Table \number\Tableno\ : #2}
\vskip 10mm
}
\def\NextTableNo{{\advance\Tableno by 1 \number\Tableno}}
\newcount\figureno     \figureno=0
\newdimen\figdim       \figdim=70mm
\def\figureinc{%
   \global\advance\figureno by 1%
}
\def\figcaption#1#2#3{\hbox to #2{\hss{\vbox{\hsize=#2 \parindent=0pt 
        {\bf Figure \number\figureno#3 :\ }#1}}\hss}
}

\def\OneSizedFig#1#2#3{\vskip 5mm
\centerline{\figureinc
  \vbox{\epsfxsize=#3\epsfysize=#3\epsfbox{#1}\vskip 5mm\figcaption{#2}{#3}{} }
  }\vskip 5mm
}
\def\OneFig#1#2{\vskip 5mm\figdim=70mm
\centerline{\figureinc
  \vbox{\epsfxsize=6cm\epsfysize=6cm\epsfbox{#1}\vskip 5mm
        \figcaption{#2}{\figdim}{}}
  }\vskip 5mm
}
\def\TwoFigs#1#2#3#4{\vskip 5mm\figdim=70mm
 \hbox to \hsize {
  \vbox {\figureinc\epsfxsize=7cm\epsfysize=7cm\epsfbox{#1}\vskip 5mm
         \figcaption{#2}{\figdim}{}
  }\hfill
  \vbox {\figureinc\epsfxsize=7cm\epsfysize=7cm\epsfbox{#3}\vskip 5mm
         \figcaption{#4}{\figdim}{}}}
 \vskip 5mm
}
\def\TwoFigsAB#1#2#3#4{\vskip 5mm\figdim=70mm
 \hbox to \hsize {\figureinc
  \vbox {\epsfxsize=7cm\epsfysize=7cm\epsfbox{#1}\vskip 5mm
         \figcaption{#2}{\figdim}{.a}
  }\hfill
  \vbox {\epsfxsize=7cm\epsfysize=7cm\epsfbox{#3}\vskip 5mm
         \figcaption{#4}{\figdim}{.b}}}
 \vskip 5mm
}
\def\FourFigsAD#1#2#3#4#5#6#7#8{\vskip 5mm\figdim=70mm\figureinc
 \hbox to \hsize {
  \vbox {\epsfxsize=7cm\epsfysize=7cm\epsfbox{#1}\vskip 5mm
         \figcaption{#2}{\figdim}{.a}
  }\hfill
  \vbox {\epsfxsize=7cm\epsfysize=7cm\epsfbox{#3}\vskip 5mm
         \figcaption{#4}{\figdim}{.b}}}
 \vskip 5mm
 \hbox to \hsize {
  \vbox {\epsfxsize=7cm\epsfysize=7cm\epsfbox{#5}\vskip 5mm
         \figcaption{#6}{\figdim}{.c}
  }\hfill
  \vbox {\epsfxsize=7cm\epsfysize=7cm\epsfbox{#7}\vskip 5mm
         \figcaption{#8}{\figdim}{.d}}}
 \vskip 5mm
}

\def\ie{\hbox{\it i.e.\ }}

\def\Sin{\hbox{sin}}
\def\Cos{\hbox{cos}}
\def\Exp{\hbox{exp}}

\def\Tan{\hbox{tan}}

\def\Atan{\hbox{atan}}

\def\C {{\rlap{\kern 1.0mm \vrule height 7pt depth 0pt} \rm C}}
\def\R {{\rlap{\kern 0.1mm \vrule height 7pt depth 0pt} \rm R}}
\def\U {{\rlap{\kern 1.2mm \vrule height 7pt depth 0pt} \rm 1}}
\debugfalse
\backetedreftrue
\eqnoperchaptertrue
\pareqreftrue 
\pareqnotrue
\eqnoperchaptertrue
\twelve
\rm
\input epsf
\def\tablecap#1{\vskip 3mm \centerline{#1}\vskip 5mm}

\def\rBBSKOne{B.M.A.G. Piette, B.J. Schroers and W.J. Zakrzewski, 
Z. Phys. C {\bf65} (1995) 165}
\def\rBBSKTwo{B.M.A.G. Piette, B.J. Schroers and W.J. Zakrzewski, 
Nucl. Phys. B {\bf 439} (1995) 205}
\def\rTigran{B.M.A.G. Piette, H.J.W. Muller-Kirsten, D.H. Tchrakian and 
W.J. Zakrzewski, Phys.Lett. B {\bf 320} (1994) 294}
\def\rSkyrme{T.H.R. Skyrme, Proc. Roy. Soc. A260 (1961) 125; G.S. Adkins, 
C.R. Nappi and E. Witten, Nucl. Phys. B{\bf 228} (1983) 552}
\def\rMichel{M. Peyrard ,B. Piette and W.J. Zakrzewski, Nonlinearity {\bf 5}, 
585 (1992).}
\def\rSasha{A. Kudryavtsev, B. Piette, W.J. Zakrzewski, Z. Phys. C 
{\bf C 60}, (1993) 731.}



\def\rLPZ{R.A. Leese, M. Peyrard and W.J. Zakrzewski,
Nonlinearity {\bf 3} (1990) 773}

\def\rChrLom{P.L. Christiansen and P. Lomdahl,
Physica 2D (1981) 482}

\def\rTDsinG{B. Piette and W.J. Zakrzewski,
in preparation }

\def\RClarkson{see {\it eg} M.J. Ablowitz and P.A. Clarkson - Solitons,
Nonlinear Evolution Equations and Inverse Scattering - CUP (1991)}
\pubnum{96/17}

\title{Mesons, Baryons and Waves in the Baby Skyrmion Model}
\authors{A. Kudryavtsev$\sp{1}$}
\authors{B. Piette,}
\authors{and}
\authors{W.J. Zakrzewski}
\DURHAM
\centerline{$\sp{1}$ also at ITEP, Moscow, Russia}
\Email{Kudryavtsev@vitep5.itep.ru \quad B.M.A.G.Piette@uk.ac.durham\quad W.J.Zakrzewski@uk.ac.durham}
 
\abstract{We study various classical solutions of the baby-Skyrmion 
model in $(2+1)$ dimensions. We point out the existence
of higher energy states interpret them as resonances
of Skyrmions and anti-Skyrmions and study their decays. 
Most of the discussion involves a highly exited Skyrmion-like state with 
winding number one which decays into an ordinary Skyrmion and a 
Skyrmion-anti-Skyrmion pair. 
We also study wave-like solutions of  the 
model and show that some of such solutions can be constructed from the 
solutions of the sine-Gordon equation. We also show that the baby-Skyrmion
has non-topological stationary solutions.  
We study their interactions with Skyrmions.
}

\chapter{Introduction.}
In previous papers by two of us (BP and WJZ) %
\REF\RBBSKTwo{\rBBSKTwo} \REF\RBBSKOne{\rBBSKOne} \REF\RTigran{\rTigran}%
[\RBBSKTwo-\RTigran]
some hedgehog-like solutions of the so called
baby-Skyrmion model were studied. It was shown there that the model
has soliton-like topologically stable static solutions (called baby-Skyrmions) 
and that
these solitons can form bound states.
The interaction between
the solitons was studied in detail and it was shown 
that the long distance force between 2 baby-Skyrmions 
depends on their relative orientation.

To construct these soliton solutions, one must use a radially symmetric ansatz
(hedgehog configuration) and reduce the equation for the soliton to an ordinary 
differential equation. This equation admits more solutions than those
described in [\RBBSKOne] and [\RTigran], and, as we will show, they
correspond to exited states or resonances made out of both Skyrmions
and anti-Skyrmions. We will also show that the model admits some
solutions in the form of non-linear waves.

The $(2+1)$-dimensional baby-Skyrmion field theory model is described by 
the Lagrangian density
 
$$ 
L=F_{\pi}\bigl({{1\over2}\partial_{\alpha}\vec \phi \partial^{\alpha} 
\vec \phi - 
{k^{2}\over4}
(\partial_ {\alpha} \vec \phi \times \partial_{\beta}\vec \phi )
(\partial^{\alpha} \vec \phi \times \partial_{\beta} \vec \phi)-
 \mu^{2}(1-\vec n \vec \phi)\bigr)}.
\eqn\eLagPhi
$$      

 Here $\vec \phi \equiv (\phi_{1},\phi_{2},\phi_{3}) $ denotes a triplet 
of scalar real 
fields which satisfy the constraint ${\vec \phi}^{2}=1$;
$( \partial_{\alpha} \partial^{\alpha}=\partial_{t}\partial^{t}-\partial_{i} 
\partial^{i})$.
As mentioned in [\RBBSKTwo-\RTigran] the first term in \eLagPhi is 
the familiar Lagrangian density of the pure
$S\sp2$ $\sigma$ model. The second term, 
fourth order in derivatives, is the (2+1) dimensional
analogue of the Skyrme-term of the 
three-dimensional Skyrme-model \Ref\RSkyrme{\rSkyrme}.
The last term is often referred to  as 
a potential term.  
The last two terms in the Lagrangian \eLagPhi
are added to guarantee the stability of a Skyrmion\Ref\RLPZ{\rLPZ}.

The vector $\vec n =(0,0,1)$ and hence the potential term 
violates the $O(3)$-rotational iso-invariance of the theory. The state 
$\phi_{3} \equiv 1$ is the vacuum state of the theory. 
As in [\RBBSKTwo,\RBBSKOne] we fix our 
units of energy and length by setting $F_{\pi}=k=1$ and choose 
$\mu^{2}=0.1$ for our numerical calculations. 
The choice $\mu\sp2=0.1$ sets the scale of the 
energy distribution for a basic Skyrmion.

As usual we are interested mainly in field configurations for which 
the potential energy at infinity vanishes as only they can
describe field configurations with finite total energy.
 Therefore we look for solutions of the equation of motion 
for the field $\vec \phi $ which satisfies

$$ 
lim_{|x|\to \infty} \vec  \phi(\vec x,t)=\vec n 
\eqn\eAssymptotic
$$                
for all $t$. This condition formally compactifies the
 physical space to a 
2-sphere $S^2_{ph}$ and so all maps from
$S^2_{ph}$ to $S^{2}_{iso}$ are characterised by 
the integer-valued degree of this map (the topological charge).
  
The analytical formula for this degree is
$$
deg[\vec \phi ]={1\over {4 \pi}} \int \vec \phi\cdot (\partial_{1} \vec \phi 
\times \partial_{2} \vec \phi) d^{2}x .
\eqn\eTopCharge
$$

This degree is a homotopy invariant of the field 
$\vec \phi  $ and so it is conserved during the time evolution.

The Euler-Lagrange equation for the Lagrangian $L$ \eLagPhi is 
  
$$
\partial^{\alpha}(\vec \phi \times \partial_{\alpha} \vec \phi-
\partial_{\beta} \vec  \phi\,(\partial^{\beta} \vec \phi \cdot \vec 
\phi \times \partial_{\alpha} \vec \phi))={\mu}^2 \vec \phi \times \vec n. 
\eqn\ePhiELeqn
$$

One simple solution of \ePhiELeqn is given by $\vec \phi(\vec x,t)= \vec n$.
This solution is of degree zero and 
describes the vacuum configuration. Another simple 
solution of (1.4) is evidently $\vec \phi=-\vec n$. It is also 
of  degree zero and may 
be considered as a false vacuum configuration.
                                     
\chapter{Static Skyrmions solutions}

Some static solutions of the equation of motion \eTopCharge were discussed 
in [\RBBSKTwo,\RBBSKOne].
An important class of static solutions of the equation 
of motion consists of fields which are invariant under the group of 
simultaneous spatial rotations by an angle $\alpha \in [0,2 \pi]$ and 
iso-rotations by $-n \alpha $, where $n$ is a non-zero integer. Such fields 
are of the form 
   
$$ 
\vec \phi(\vec x)= \left(
\matrix{ &\sin f(r) \cos (n\theta)\cr
 &\sin f(r) \sin(n\theta)\cr
 &\cos f (r)\cr}
\right),
\eqn\eHedgehog
$$                         
where $(r,\theta)$ are polar coordinates in the $(x,y)$-plane and $f$ is a 
function which satisfies certain
 boundary conditions which will be specified below. Such 
fields are  analogues of the hedgehog field of the Skyrme model and were 
 studied in  [\RTigran] for different values of $\mu^2$.

The function $f(r)$, the analogue of the profile function
of the Skyrme model, has to satisfy 
$$ 
f(0)=m\pi ,  m \in Z
\eqn\efzero
$$                                  
for the field \eHedgehog\ to be regular at the origin. To satisfy the boundary 
condition \eAssymptotic\ we set
 
$$ 
lim_{r \to \infty}f(r)=0. 
\eqn\efinfinity
$$
Then solutions of the equation of motion which satisfy
 \efzero and \efinfinity\ 
describe fields for which the total energy is finite. 
Moreover, the degree of the fields (2.1) is
   
$$        
deg[\vec \phi]=[\cos f(\infty)-\cos f(0)]{n\over 2}. 
\eqn\eDegree
$$

For fields which satisfy the boundary conditions \efzero\ and 
\efinfinity\ and which thus
correspond to finite energy configurations, we get from \eDegree

$$ 
deg[\vec \phi]=[1-(-1)^m]{n\over 2}. 
\eqn\eDegreeMN
$$

The fields of the form \eHedgehog which are stationary points of the static 
energy functional $V$, the time independent
part of $L$ in (1.1), must satisfy the Euler-Lagrange equation for $f$

$$ (r+{n^{2}\sin^{2}f\over r})f^{\prime\prime}+
(1-{n^{2}\sin\sp2 f \over r^{2}} +{ n^{2}f^\prime \sin f\cos f \over r})f^\prime-
{n^2\sin f\cos f \over r} -r{\mu}^{2}\sin f=0.
\eqn\efeqn
$$
   
In [1,2] it was shown that the solutions of \efeqn for $n=1,2$ and 
$m=1$, correspond to the absolute minima of the energy functional 
of degree $1$ and $2$ respectively. 
Any field obtained by translating and 
iso-rotating the solution corresponding to $n=m=1$
was called a baby Skyrmion. As 
this field configuration has a topological charge $deg[\vec \phi]=1$
we call it a ``baryon" and denote it  by the symbol $B$. 
A solution corresponding to 
$n=-1$ and $m=1$ is then an anti-Skyrmion or an antibaryon $\bar B$.

Clearly, there exist also solutions of the equation of motion 
which satisfy the boundary 
conditions \efzero\ at the origin but which, at infinity, behave as
  
$$
\eqalign{
&lim_{r \to \infty } f(r)=l\pi.\cr
&l \in Z.\cr}
\eqn\efother
$$

For such fields, if $l$ is odd, the solutions \efeqn differ from 
$B$-Skyrmions and belong to a class of solutions with infinite energies. 

It is also worth mentioning that the asymptotic behaviour of these
$\infty$ energy Skyrmion-like 
solutions differs from the asymptotic behaviour 
of $B$-Skyrmions. Consider, e.g., 
the case $m=1,l=0,n=1 $ ($B$-Skyrmion).
Then the asymptotic behaviour of this solution is given by
[\RBBSKOne]:

$$ 
f(r)\sim K_{1}(\mu r) \sim_{r \to \infty} \sqrt{\pi \over 2\mu r} e^{-\mu r}
\eqn\eAsympSk$$
where $K_{1}(x)$ is the modified Bessel function. On the other hand the asymptotic behaviour of the
 $\infty$ energy anti-Skyrmion-like solution  corresponding to
$(n=1,m=2,l=1)$ is given by
     
$$ 
f(r) \approx \pi+ {C\over  \sqrt { \mu r}}\Sin(\mu r+\gamma),
\eqn\eAsymptASk$$
where $C$ and $\gamma $ are constants.
As the energy of these solutions is infinite we will not
discuss them further in this paper and so we will always assume that 
$l=0$ in what follows.

\TwoFigsAB{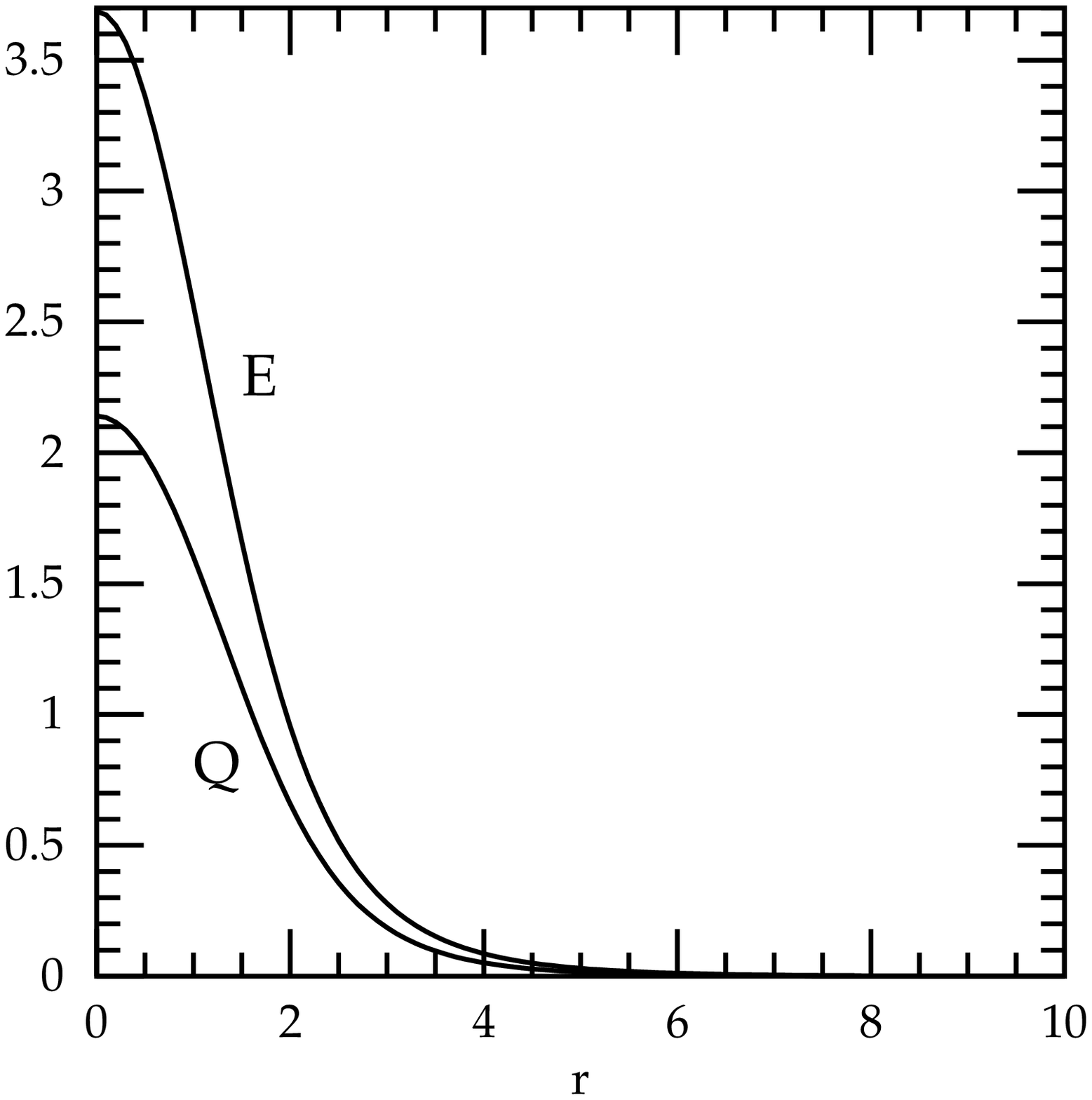}{Energy and Topological charge density for the hedgehog 
solution : n=1, m=1}{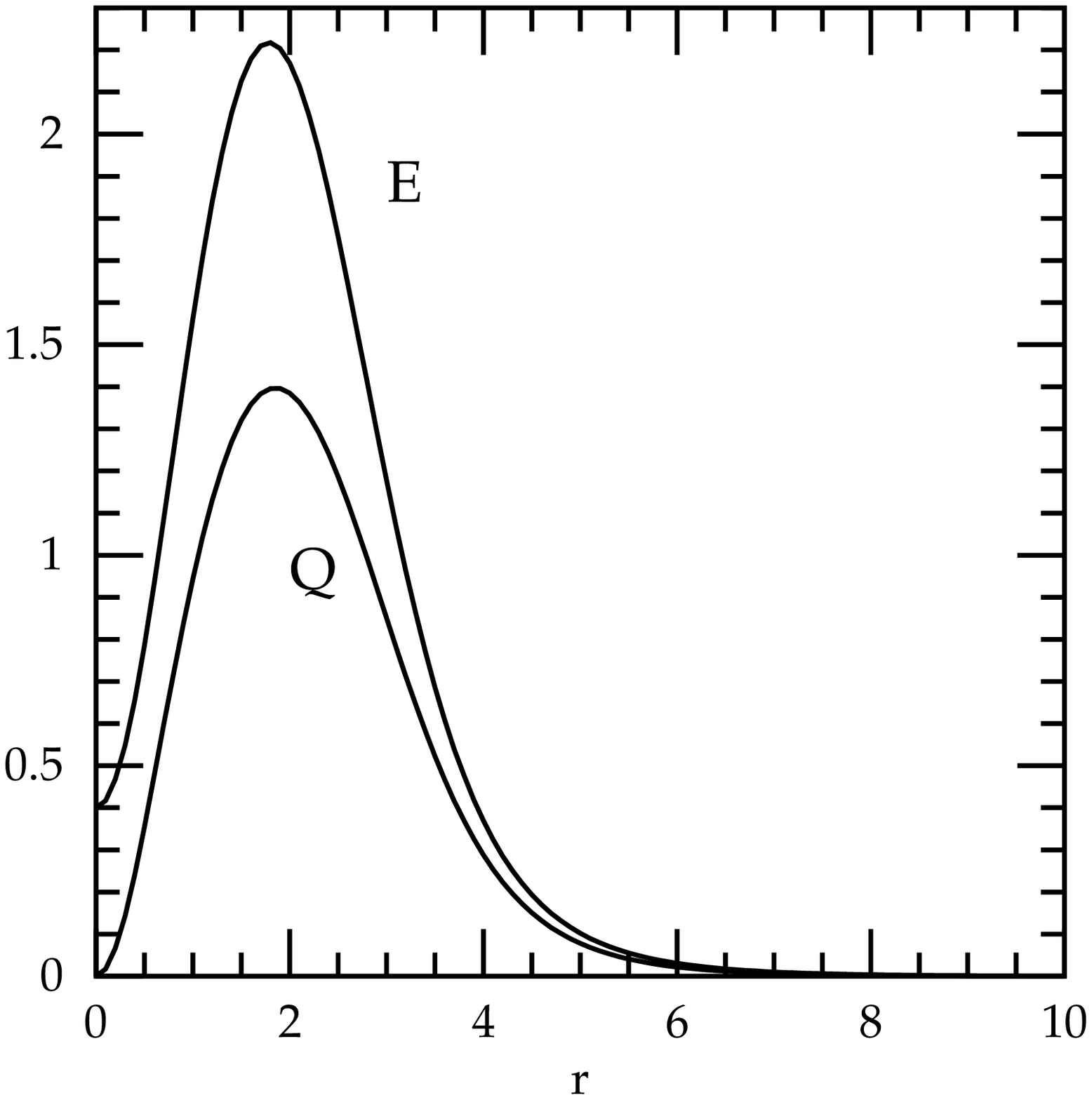}{Energy and Topological charge density for the hedgehog solution : n=2, m=1}

We have solved \efeqn\ for different values of $n$ and $m$ 
using a shooting method with the appropriate boundary condition 
\efzero \efinfinity.
We have determined the profile function $f(r)$ and the
energy and topological density profiles for each of these solutions.
The total energies for our solutions are as follows:

\vskip 3mm
\centerline{\vbox{\offinterlineskip\tabskip=0pt\halign
{\strut\vrule#&\quad#\hfill\quad&\vrule#&\hfill\quad#\quad\hfill&\vrule#
&\hfill\quad#\quad\hfill&\vrule#&\hfill\quad#\quad\hfill&\vrule#
&\hfill\quad#\quad\hfill&\vrule#\cr
\noalign{\hrule}
&n\quad$\backslash$\quad m &&1 &&2 &&3 &&4&\cr
\noalign{\hrule}
&1 && 1.5642 && 5.011 && 10.030 && 16.492&\cr
&2 && 2.9359 && 7.725 && 14.185 && 22.233&\cr
&3 && 4.4698 && 10.555 && 18.350 && 27.819&\cr
&4 && 6.1145 && 13.465 && 22.633 && 33.395&\cr
\noalign{\hrule}
}}}
\vskip 3mm
When $m=1$, the topological charge is given by $n$, and each configuration
is a superposition of $n$ Skyrmions. 
We know from [\RBBSKOne] that only the first two are stable. 
Figure 1, shows the profiles of the energy and of the topological density
of the states $n=1$ and $n=2$.

When $m=2$, the topological charge is 0 and, as can be seen from
Figure 2, these configurations correspond 
to a superposition of $n$ Skyrmions and $n$ anti-Skyrmions where the
Skyrmions form a ring surrounding the anti-Skyrmion at the centre. 
We have integrated separately both the positive and negative parts of 
the topological charge, and have found them to be $n$ and $-n$
respectively, thus justifying our interpretation.

\TwoFigsAB{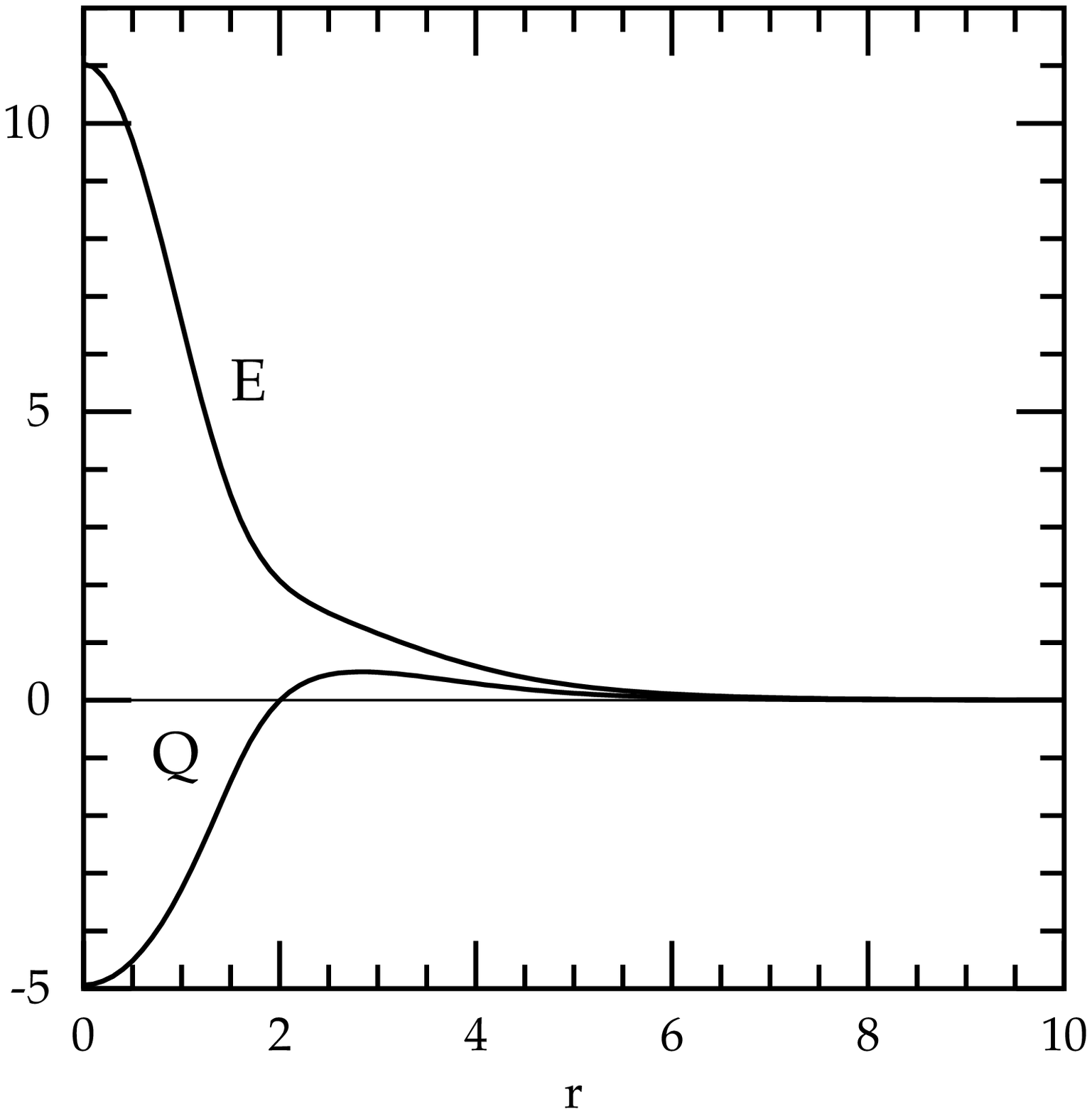}{Energy and Topological charge density for the hedgehog 
solution : n=1, m=2}{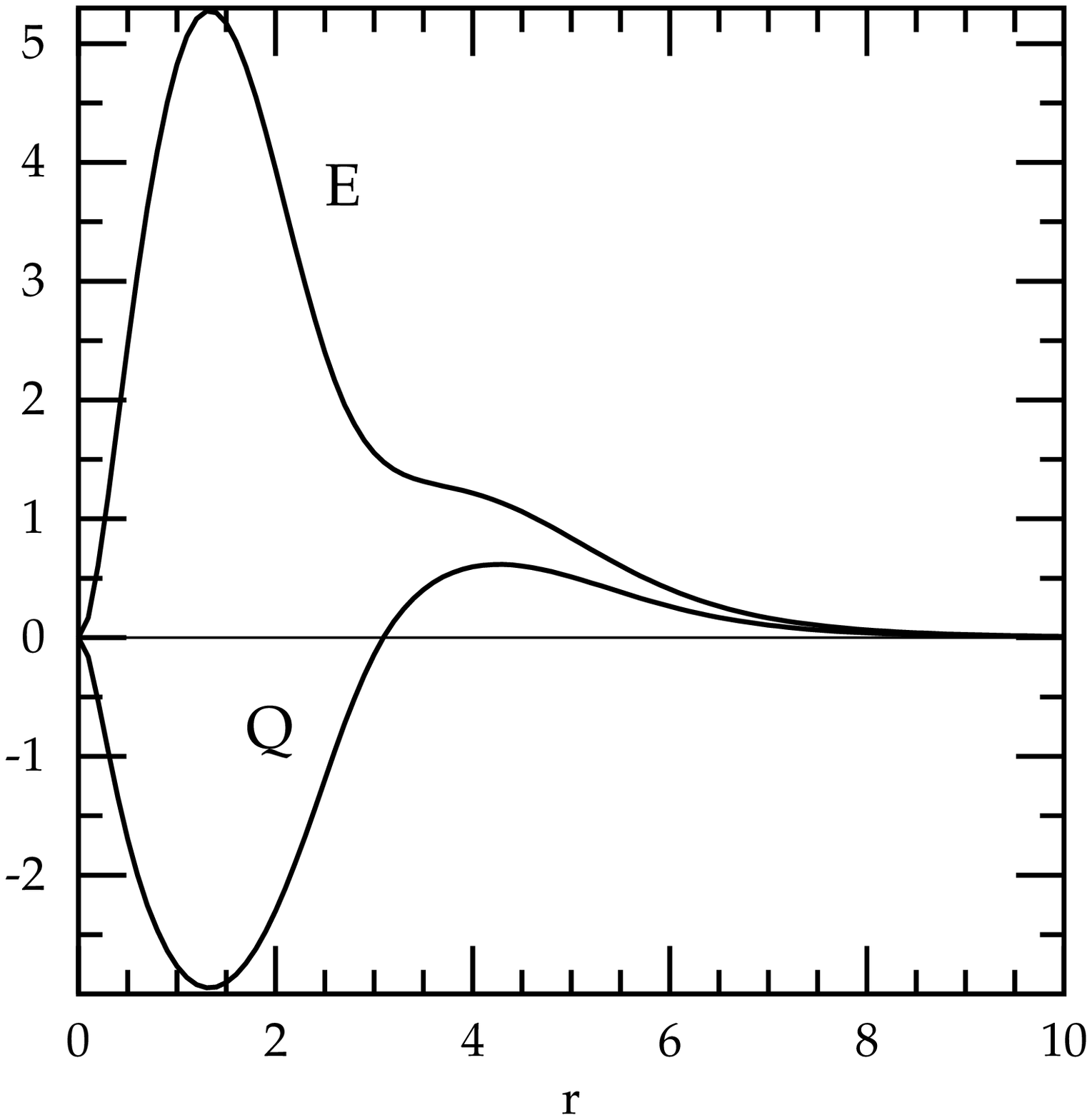}{Energy and Topological charge density for the hedgehog solution : n=2, m=2}
  
When $m=3$, the topological charge is given by $n$, and each configuration
is a superposition of $2n$ Skyrmions and $n$ anti-Skyrmions (Fig. 3).
The configuration is made of 3 layers, with $n$ Skyrmions at the centre,
$n$ anti-Skyrmions in the middle, and $n$ Skyrmions in the outside ring.  

\TwoFigsAB{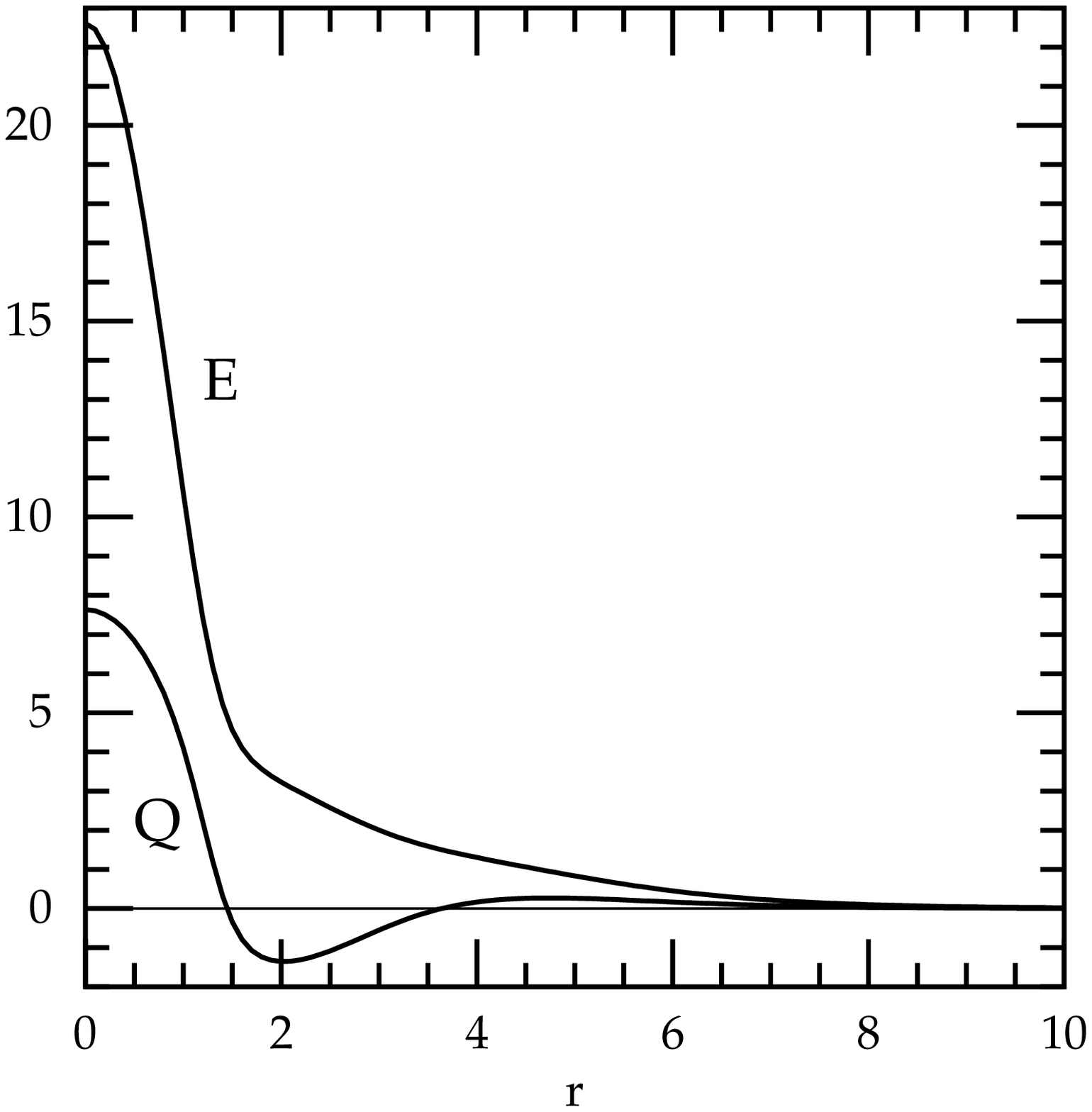}{Energy and Topological charge density for the hedgehog 
solution : n=1, m=3}{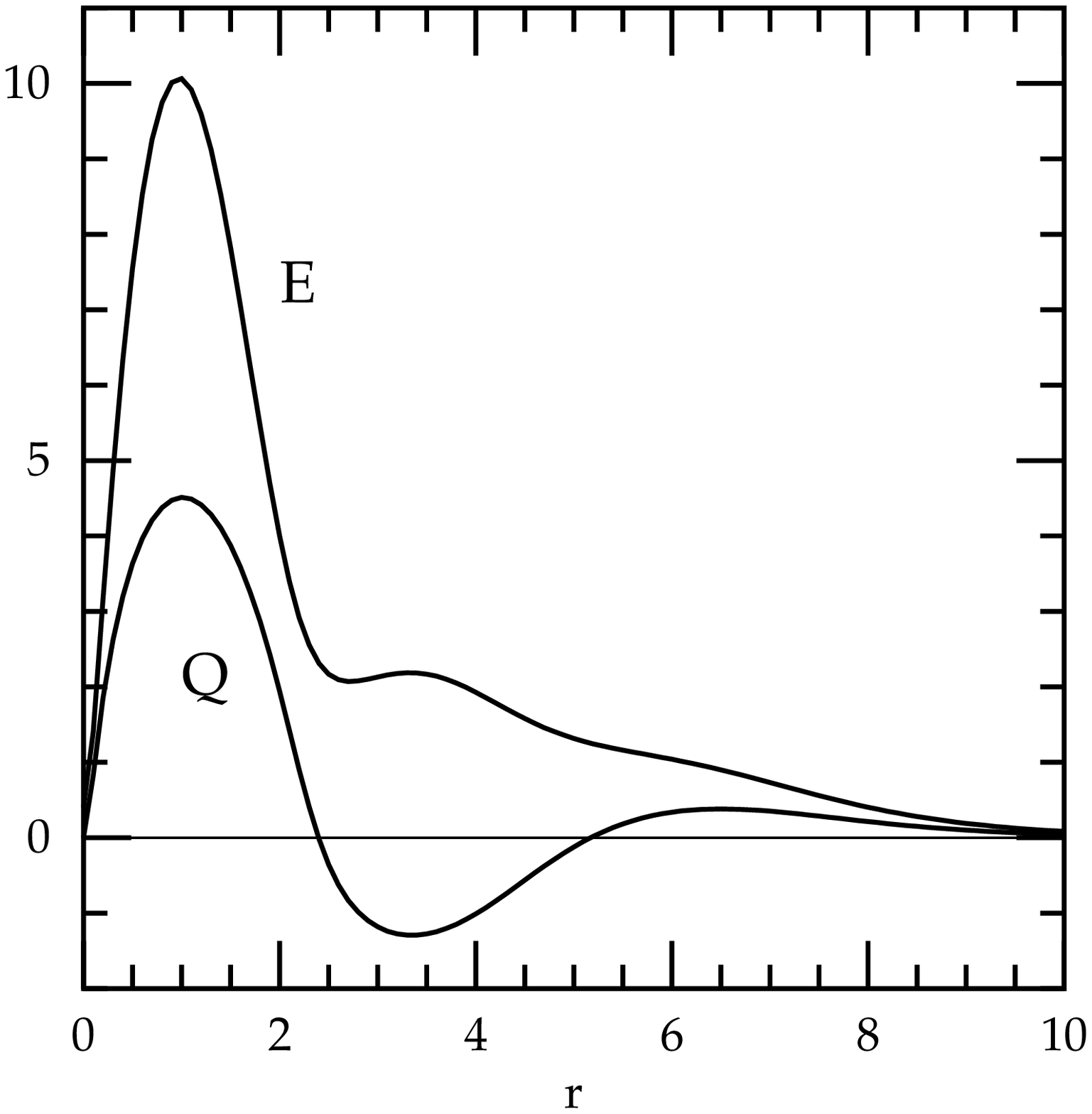}{Energy and Topological charge density for the hedgehog solution : n=2, m=3}

Note that when $m$ is larger than 1 the energy of the
configuration is larger than the energy of $n\,m$ baby Skyrmions. 
This indicates that such configurations are unstable and we will now analyse
their decay modes.

\chapter{Exited meson $ M_{1,2} (n=1;m=2)$}
Let us look first at the field which corresponds to $n=1,\;m=2$.
From (2.4) we see that the topological charge of this field 
configuration is zero. So we can call this state,
a ``meson" or a coherent ``meson cloud".
We note from Figure 2.a that, like a $B$ Skyrmion, this solution corresponds 
to a radially symmetric extended configuration with the maximum of
the energy density at the origin $(r=0)$.
 
As the total energy ($5.011$) exceeds the sum of the masses of a $B$-Skyrmion 
and a $\bar B$-Skyrmion (2 x 1.5642), this configuration must be unstable.
 
 The distribution of the topological charge density shows more
structure (Fig 2.a.)
We note that the topological charge density
is negative for small $r$ and that it changes sign at $r=r_{cr}=2$. 
At this point the profile function 
$f(r_{cr})=\pi$, thus the solution corresponding to
the $M_{1,2}$ state looks like a Skyrmion surrounded by an anti-Skyrmion
field. Of course the total topological charge is zero.
We see that the solution  still looks like an 
extended but localised configuration. However, although the topological
charge density suggests that the Skyrmion is at the origin 
the profile function there 
is given by $f=2\pi$ and so, from this point of view,
 resembles more the vacuum than a Skyrmion.
This suggests a possible interpretation of the $M_{1,2}$ state
in terms of Skyrmions and anti-Skyrmions of the usual or 
$\infty$ energy type. Whatever the interpretation, the state
is mesonic in  nature. Moreover, looking at the field configuration
we note that most of its changes takes place around those points
in the $(x,y)$ plane where $r=r_{cr}$ {\it i.e.} where $f=\pi$.
It is remarkable that everywhere along this circle 
$\phi_{3}=-1$. This circle is actually the region of instability when the 
solution $M_{1,2}$ is excited by a non-radial perturbation of small amplitude.

In fact, it is not difficult to demonstrate that at each point on the 
circle $r=r_{cr}$ one can create  ``hedgehog-like" extended objects  
using only infinitesimal perturbations.
Of course, if these extended objects were to correspond 
to Skyrmions and anti-Skyrmions then to  conserve the topological charge 
they will have to be created in pairs and we would expect them to 
appear as soon as we perturb the initial configuration.
To check for such a behaviour we have perturbed the initial
configuration corresponding to the  $M_{1,2}$-state. Our 
perturbation was in the form of small excess of kinetic energy centered around 
two points in the (x,y)-plane chosen symmetrically with respect to 
$r=0$. 
We have found that, indeed, the state split into
a Skyrmion and anti-Skyrmion pair of the 
$B$-type.  Their relative orientation was
such that the force between them was repulsive. After their creation the 
Skyrmion and the anti-Skyrmion moved in 
opposite directions from the centre.
  
We then experimented with perturbing the initial state by different
perturbations and we observed different decay modes.  
When the applied perturbation was not symmetric with respect to $r=0$ 
(but was close to being symmetrical)
the $M_{1,2}$ state decayed into a Skyrmion and an
anti-Skyrmion, which then rotated in their internal space so that their
relative orientation made them to attract each other. 
They then collided into each other and decayed into waves.
   
We thus conclude  that the solution $M_{1,2}$ of  
(2.6) is indeed a saddle point in the space of field configurations.
The state can be thought of as a resonance
of a Skyrmion and an anti-Skyrmion and it has different decay modes
when perturbed. It decays either  
into a $B\bar B$-pair or into light ``mesons".

It would be interesting to see whether it is
possible to create the $M_{1,2}$ state in a head-on collision 
of a Skyrmion and an anti-Skyrmion.  We are planning to come back to this 
question in a future paper.

\chapter{Exited $B_{1,3}$ baryon $(n=1,m=3)$}

This is a baryon-type state as its topological charge  is one. The 
profiles of the energy and of the topological charge densities  
for this state are shown in Figure 3a.
Looking at the density of the topological charge
we see that this state may be considered as the 
usual $B$-Skyrmion surrounded by an anti-Skyrmion and a further Skyrmion ring.
The configuration has an energy larger than the energy of its constituents
and is thus unstable. 

We have performed several simulations looking at the decay 
products of this state. When we used a perturbation symmetric with respect
to the origin the configuration decayed in 2 Skyrmions and 1 anti-Skyrmion,
the anti-Skyrmion staying at the origin, while the two Skyrmions moved in
opposite directions. When the perturbation was not symmetric, the
Skyrmions and the anti-Skyrmion were able to change their relative orientations 
and one of the Skyrmions collided with the anti-Skyrmion and decayed into waves.

\chapter{Mesons made from dibaryons and anti-dibaryons}

Let us now discuss various properties and decay modes of exited
meson-like states made out of dibaryons (a bound state of 2 $B$-Skyrmions)
and antidibaryons.
We will concentrate our attention  on the  
$n=2$ and $m=2$ state but our discussion
generalises easily to other states. The topological charge 
of the ($n=2$, $m=2$) configuration ({\it ie} its baryon number) 
is zero, so this configuration is a meson-like state from 
the point of view of our classification. As $n=2$ the configuration looks 
like a dibaryon near the origin, i.e. near  $r=0$.
From Figure 2.b, we see that it corresponds to a dibaryon at the origin
surrounded by an anti-dibaryon ring. 
The border between these two regions of opposite topological charge is, again,
very well defined and  
is situated along the circle of radius $r_{cr} = 3$, ($f(r_{cr})=\pi$). So 
this ring is again the region of instability. Applying different types of 
perturbations to our  ($n=2$, $m=2$) solution, we have observed  three 
different decay modes for this state:

$$ M_{2,2}\rightarrow \cases{  B+\bar B+B+\bar B&\cr  B+\bar B+waves&\cr waves&\cr}.
\eqn\eBttDecay$$
 
The pictures of the energy density for the first decay modes is 
shown in Figure 4.

\OneFig{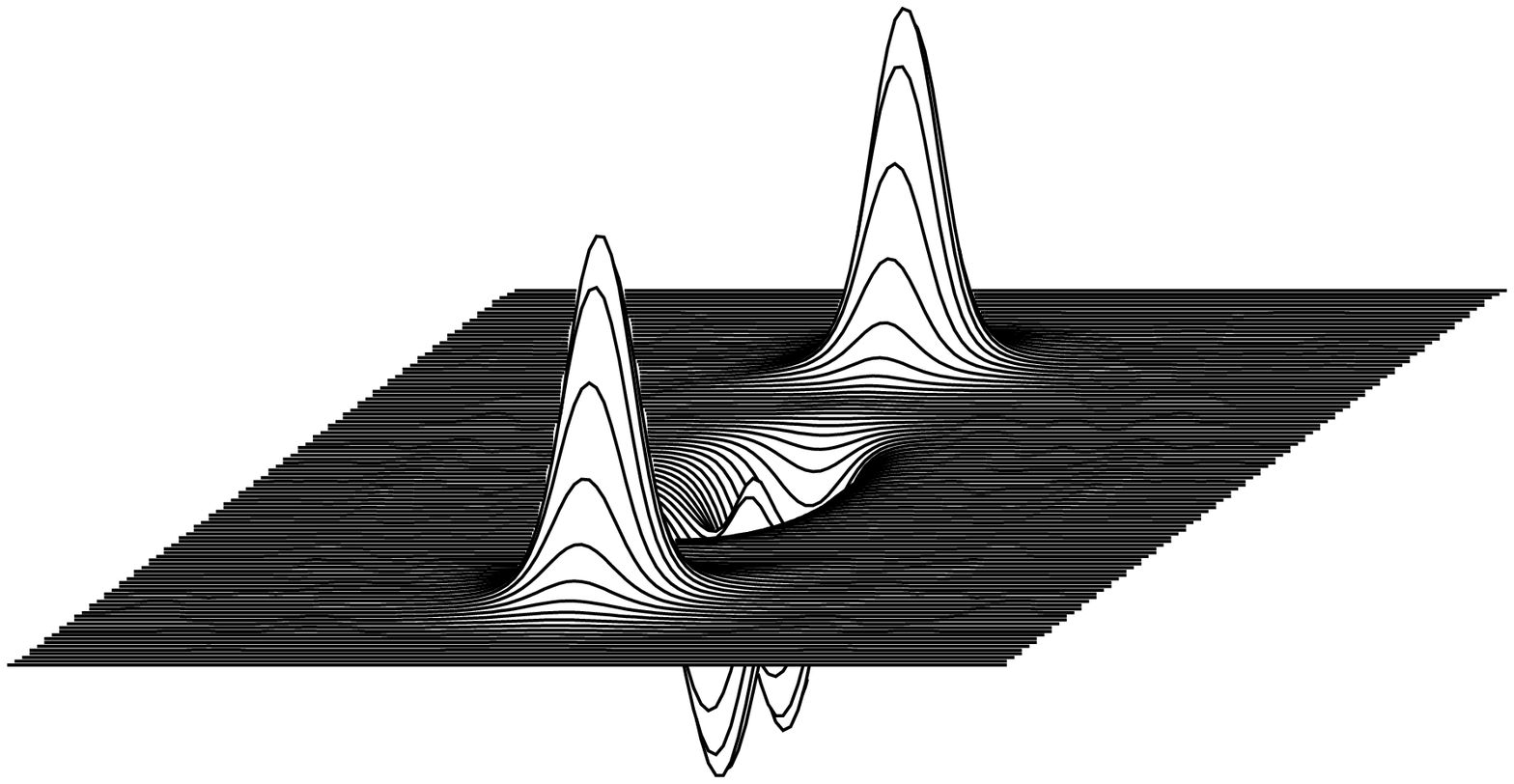}{Topological charge density for the 
$M_{2,2}\rightarrow  B+\bar B+B+\bar B$ decay mode.}

\chapter{Other exited mesonic and baryonic states}
  
\item{i)}Let us look first at the state $(n=1,m=4)$. This state, again, is
mesonic with $deg[\vec \phi]=0$.
The energy and topological charge profiles are shown in Figure 5. 
The state is more complicated as its energy density exhibits 
additional maxima and minima. In fact,  the state looks 
as if it were a coherent state of Skyrmions and 
anti-Skyrmions, with rings of different radia occupied by fields of 
alternating topological charge. This is clearly seen from the topological
charge density plots; moreover, the total topological charge 
in each ring is +1,-1,+1 and -1 respectively (going out from $r=0$). 
Thus we denote this state as $M_{1,4}$ meson.

\TwoFigsAB{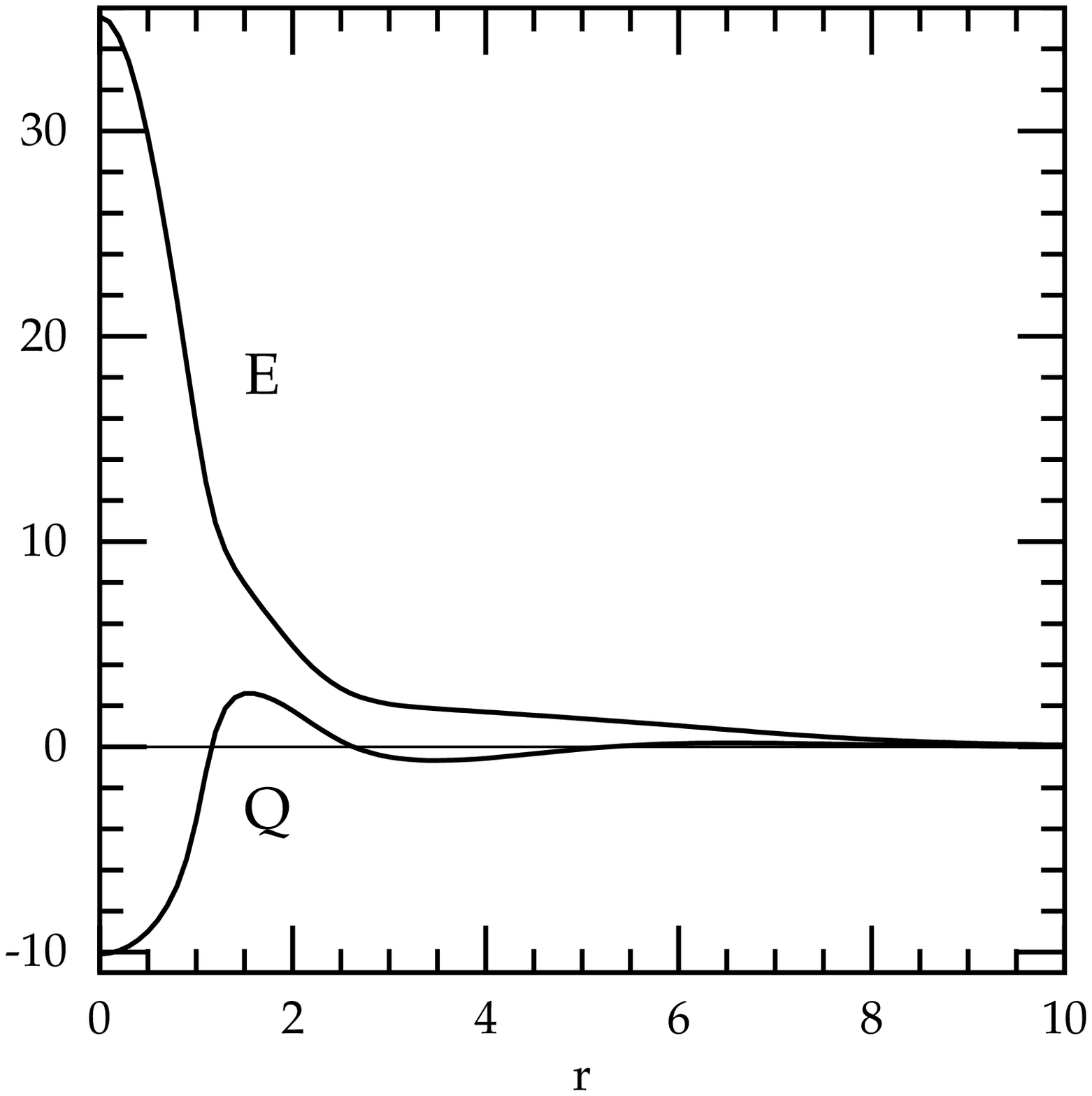}{Energy and Topological charge density for the hedgehog 
solution : n=1, m=4}{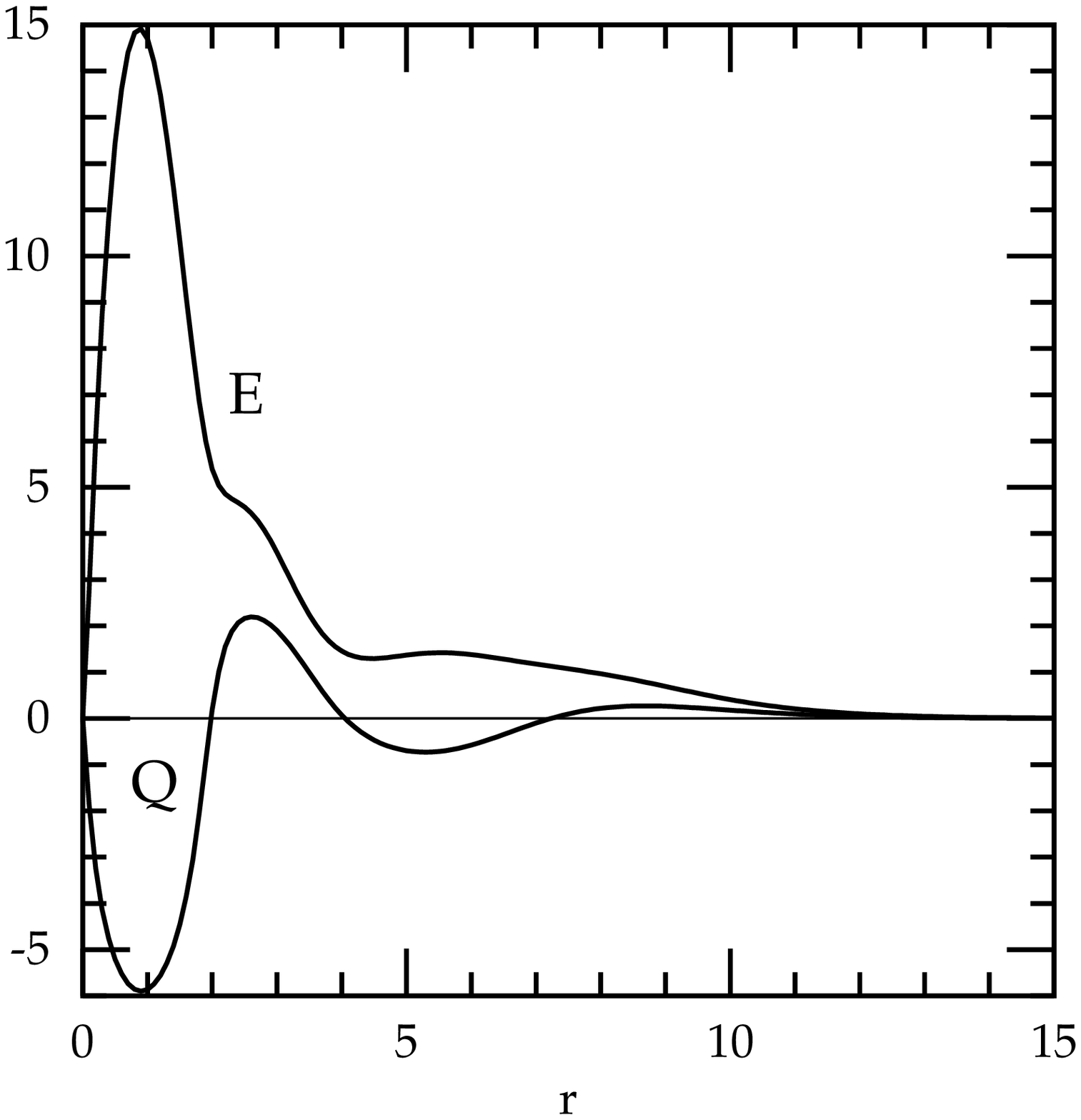}{Energy and Topological charge density for the hedgehog solution : n=2, m=4}

\OneFig{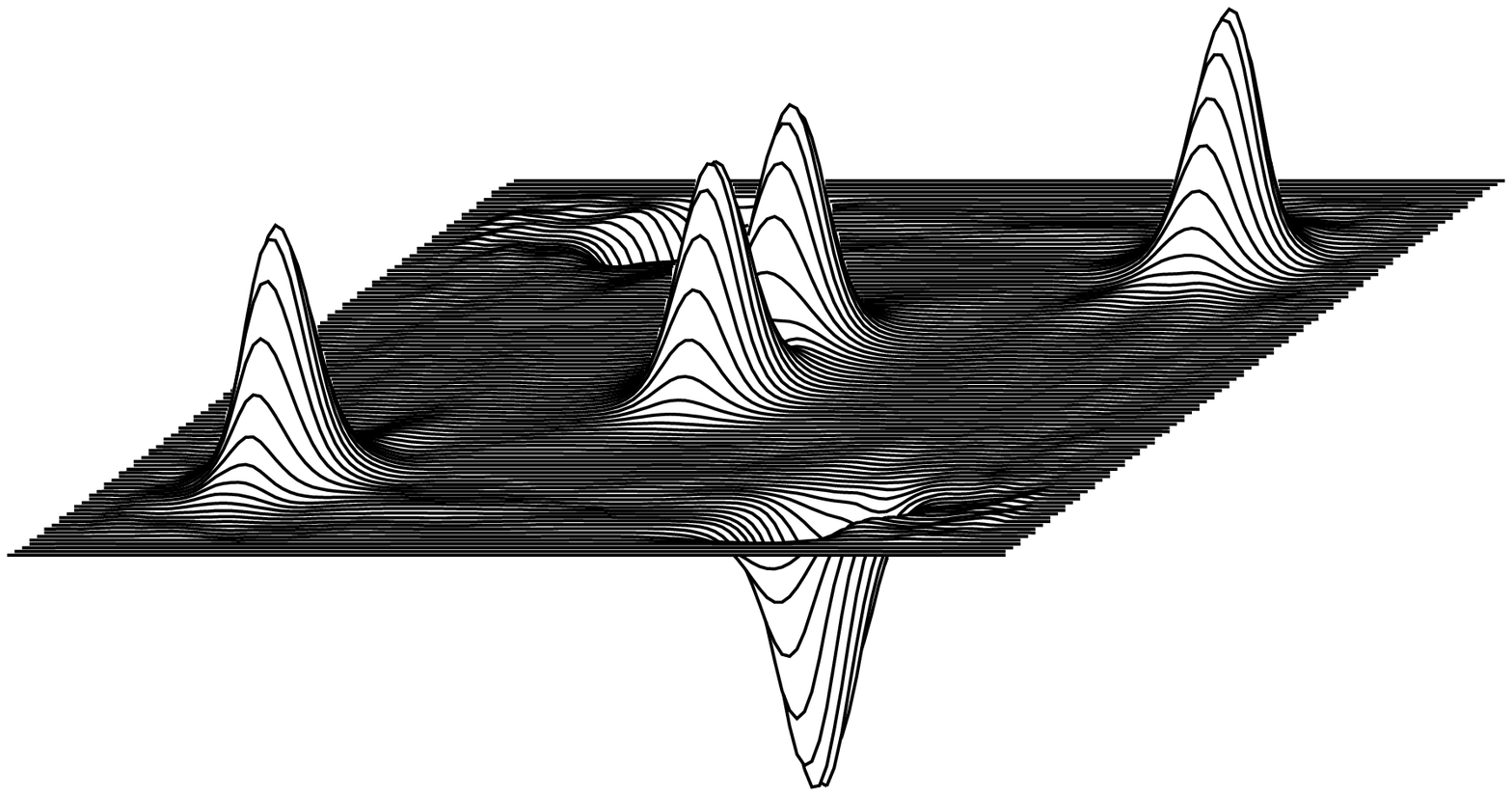}{Topological charge density for the
$M_{2,3}\rightarrow  B+\bar B+B+\bar B + B + B$ decay mode.}

\item{ii)} 
Another interesting state is that of $M_{2,3}$ $(n=2,m=3)$.
Its topological charge is $2$.
Hence this state can be thought of as an exited state of the dibaryon 
$M_{2,1}$. Its energy is clearly quite large 
and the state represents a coherent 
mixture of 4 Skyrmions and 2 anti-Syrmions and so
it can decay into different channels. 
One  example of its decay mode is shown in Figure 6. 
We see in this picture that the decay products consist of two 
outgoing Skyrmions and  two anti-Skyrmions with 
the remaining 2 Skyrmions at rest at the origin.
The fact that the decay products involve one di-Skyrmion
at the origin (slightly excited) is due to the attraction between 
two Skyrmions, as $\hbox{mass}(M_{2,1}) < 2\,\hbox{mass}(B)$, see our table
in section 2.
 

\item{iii)} Another interesting  state is $M_{2,4}$
which corresponds to $(n=2,m=4)$ and, as such, is  a 
highly excited meson state. Its 
energy and topological charge profiles are shown in Figure 5b.

Clearly, the list of new solutions may be continued further by taking 
larger values of $n$ and $m$. Of course, having calculated 
some of their profile functions we see that as one 
increases $m$ their energies increase (quite rapidly). 
All these higher energy states can be treated as coherent states
of some number of Skyrmions and anti-Skyrmions
and are unstable. Under suitable perturbations they will decay
into various channels involving Skyrmions and anti-Skyrmions with
some Skyrmions and anti-Skyrmions annihilating into pure waves.

\chapter{Plane wave solutions}
Let us now look at wave-like solutions of our model.
Recall that such  wave-like  solutions have already been
studied  in \Ref\RMichel{\rMichel} \Ref\RSasha{\rSasha} 
for a slightly different version of a ``skyrme-like" (2+1)
dimensional model (the potential term of that model was different.) 

What type of  waves can we find for the Lagrangian in the form \eLagPhi?
To look for plane wave-like solutions, we seek  solutions
which do not  depend on one variable, say  $y$. 
However, as soon as we impose this condition we note that the  
Skyrme-like term vanishes for these field
configurations. So the discussion is not that difficult
and can be performed analytically. 

First, we look for solutions of the
equation  of motion for the field $\vec \phi$ in the form:
 
$$    
\vec\phi=(\sin{f}\cos{\psi},\sin{f}\sin{\psi},\cos{f}),  
\eqn\eVecPhi
$$ 
where $f=f(x,t)$ and $\psi=\psi(x,t)$. In terms of  $f$ and $\psi$
the Lagrangian takes the form
 
$$
L=(1/2)\int d^{2}x \,{(\partial_{\mu}f\partial^{\mu}f + 
((1-\cos(2f))/2)\partial_{\mu}\psi \partial^{\mu}\psi- \mu^2 (1-\cos(f))}.  
\eqn\eLagf
$$

To go further we consider the case when the 
phase $\psi$ is constant.
In this case   the field $f$ satisfies
$$
\partial_\mu \partial^\mu f+ \mu^{2}\sin(f)=0,
\eqn\eSinGordon
$$
which is the sine-Gordon equation. 
So we see that when $\psi=const$
the wave solutions of the Lagrangian (1.1) are given by the solutions 
of the sine-Gordon equation.

Moreover, solutions of \eSinGordon with small 
amplitude may be considered as  ordinary plane waves with the dispersion 
relation given by
       
$$
\omega^{2}=\mu^{2}+k^{2}.
\eqn\eOmega
$$

Of all finite energy solutions of \eSinGordon (one-dimensional 
case) the most important, and perhaps the best studied, is the
solution of the kink type:
  
$$
f(x,t)=4\,\Atan\Bigr(\Exp\Bigr[-\mu({x-x_{0}-vt\over\sqrt(1-v^2)})\Bigl]\Bigl).
\eqn\eKink
$$

When translated to our case we note that when $x$  changes from $-\infty$ to 
$+\infty$, the vector $\vec\phi$ 
moves along the meridional
cross-section of the $S^2_{iso}$-sphere and returns to the same point.
Moreover, this is true for any fixed time $t$.

Another solution of \eSinGordon, called breather, is also well known
and has been studied by many people. Its form is\Ref\rClarkson{\RClarkson}
$$
f(x,t)=4\Atan\Bigr({(1-\omega^2)^{1/2}\over\omega}
    {\sin(\omega(t-t_0))\over\cosh((1-\omega^2)^{1/2}(x-x_0))}\Bigl)
\eqn\eWHAT                                             
$$ 
where $\omega$ is the frequency of the internal breather oscillations.

These 2 types of solitonic waves are infinite front lines (the solutions
depend only on one variable, $x$). Unfortunately, once imbedded into $S^2$,
the target space for our model, the extra degrees of freedom make these waves 
unstable. For the kink
solution this is not surprising as the ``loop" around the meridian can easily
``slip" on one side of the sphere, thus decreasing the potential energy.

To see this, let us take the configuration 
$$
\vec \phi(\vec x)= \left(
\matrix{ &\Cos \alpha \,\Sin f \cr
 &\Sin\alpha \,\Cos\alpha\ \,(1-\Cos f)\cr
 &1-\Cos^2\alpha \,(1-\Cos f)\cr}
\right),
\eqn\ePertKink
$$ 
where $f$ is given by \eKink with $v$ set to $0$, and where $\alpha$
is a function which depends only on $y$ and which goes to $0$ when
$y$ goes to infinity. This configuration describes the kink \eKink\eVecPhi
perturbed locally in $x$ and $y$, and it ``displaces" the kink from the meridian
of the 2-sphere onto a loop of a smaller radius on $S^2$. 
The energy density for this static configuration is given by
$$
\eqalign{
E = \int dx dy \Bigl[ {1\over2} \bigl[ \Cos^2\alpha f_x^2& +
    \alpha^2_y (\Sin^2\alpha \Sin^2 f + (1-\Cos f)^2)  \cr
    + k^2 f_x^2 \alpha^2_y (1-&\Cos f)^2 \Cos^2\alpha\bigr]
   + \mu^2 \Cos^2\alpha (1-\Cos f)\Bigr].\cr
}
\eqn\ePertEn
$$
To prove that for some appropriate choice of $\alpha$, the energy  
decreases, we compute the change of energy induced by a non-zero $\alpha$
$$
\eqalign{
\delta E =& E(\alpha=0) - E(\alpha)\cr
        =& \int dx dy \Bigl[2 \Sin^2{f\over 2}\bigl[\mu^2(\Sin^2\alpha-
            k^2\alpha_y^2(1-\Cos f)^2 \Cos\alpha)
            -\alpha_y^2(1+\Sin^2\alpha\Cos{f\over 2})\bigr]\cr
         &  +\mu^2\Sin^2\alpha(1-\Cos f)\Bigr]\cr
       \ge& \int dx dy \Bigl[2 \Sin^2{f\over 2}\bigl[\mu^2(\Sin^2\alpha-
            k^2\alpha_y^2 \Cos\alpha)
            -\alpha_y^2(1+\Sin^2\alpha)\bigr]
            +\mu^2\Sin^2\alpha(1-\Cos f)\Bigr]\cr
}
\eqn\eDeltaE
$$
where we have used the fact that for \eKink\ $f_x = 2 \mu sin(f/2)$.
Given $\alpha(y)$ satisfying the asymptotic behaviour imposed, we
can always stretch it by performing the dilation $y \rightarrow a y$
to make $\alpha_y$ small enough to make the two negative terms in 
\eDeltaE\ smaller than the positive terms. This proves the instability of
the sin-Gordon kink wave when imbedded into the $S^2$ model.

We have performed some numerical simulations for both types of waves and have 
indeed observed their instability. In both cases, as soon as some region of the 
wave is perturbed, the wave collapses around this point, emitting radiation. 
The collapse front then propagates rapidly along the solitonic wave 
destroying it completely.

In our previous work \refmark\RSasha, we have studied the scattering properties 
of plane waves and skyrmions. The skyrmions studied in [\RSasha] were 
different (the potential was different) so we have repeated our 
analysis for the model studied here and have found no major difference. 
Like in the previous case, the skyrmion absorbs a section of the wave 
and starts moving after the collision. 

We have also studied the scattering between a skyrmion and the sine-Gordon
front waves. We observed that when the skyrmion is close to
a sine-Gordon wave (breather or kink) it triggers the wave collapse and, as a 
result, there is no real scattering between these two objects.

While performing simulations with large amplitude breather waves we have
observed the formation of a radially symmetric breather-like soliton. 
This solution looks very similar to the pulsons observed in
\REF\RChrLom{\rChrLom}[\RChrLom]. We discuss some of its properties
in the next section.

\chapter{Non-topological Solitons}
In the previous section we have described our studies of waves in the 
baby-skyrmion model in which we observed that the plane wave solutions which 
are given by the solutions of the (1+1) dimensional sin-Gordon equation are 
unstable.

However, we when we looked at the decay of some breather front
 waves we encountered something rather unexpected. Instead of 
decaying into waves that dissipate like kink solutions they produced a
radially  symmetric breather-like field configuration
 which appeared to be relatively 
stable. By looking at the time evolution of this field configuration, 
as produced by our simulation,
we were able to conclude that:
\item{-} the field configuration is radially symmetric.
\item{-} up to a global rotation of $S^2$, the solution ``lives'' in the
$\phi_1,\phi_3$ plane of the target space ($S^2$).

To study this field configuration further we make the following ansatz:
$$    
\vec\phi=(\Sin{f(r,t)},0,\Cos{f(r,t)}).  
\eqn\eVecBreather
$$
The lagrangian density then becomes 
$$
L=\pi\int dr\,r\,{(\partial_{t}f\partial_{t}f - 
\partial_{r}f\partial_{r}f - \mu^2 (1-\Cos(f))}.  
\eqn\eLagf
$$
and the equation reduces to the radial sin-Gordon equation:
$$
f_{tt} - f_{rr} - {f_{r}\over r} + \mu^{2}\Sin(f)=0.
\eqn\eSinGordon
$$

This equation has already been studied in [\RChrLom], where it
was shown that it has time dependent solutions similar to a breather, but
which radiate their energy and slowly die out. The authors 
of  [\RChrLom] decided to call 
such configurations  pulsons.
 
In \REF\RTDsinG{\rTDsinG}[\RTDsinG] we have also shown 
that there exist stable time
dependent solutions. The radial field configurations of \eSinGordon\ radiate
relatively quickly when their amplitude of oscillation is relatively small, 
that is when the value of $f$ never becomes larger than $\pi/2$ at the origin 
(these are the pulsons studied in [\RChrLom].) 
When the amplitude of oscillation is larger than  $\pi/2$ the breather-like
configuration slowly radiates its energy and asymptotically reduces 
the amplitude of 
oscillation to $\pi/2$ and and settles at a period of oscillation $T \sim 20.5$ 
(when $\mu^2= 0.1$). By trial and error we have found that
$$
\eqalign{
f(r,0) = 4\, \Atan\Bigl(C \,\Exp\bigl(-{2\over\pi}& {\mu r\over K} \Atan({\mu r\over K})\bigr) \Bigr)\cr
{\partial f \over \partial t} (r,0) &= 0 \cr
}
\eqn\ePBinit
$$
with $K = 10$ and $C= \Tan(\pi/8)$ is a good initial condition for this 
metastable pseudo-breather solution.

As in the case of  plane wave solutions
we have to check that once they are imbedded into $S\sp2$ they
are still stable. We have indeed checked this numerically and we have found
that the solution described by \eVecBreather\ and \eSinGordon\ is indeed 
stable in the $S^2 $ model. 

The energy of this new breather-like solution is given by
$$
  E_{PB}\, \sim \,3.97
$$
which means that it is $2.5$ heavier than the baby-skyrmion.
Moreover, its topological charge density is identically zero
 but  it has enough energy
to decay into a skyrmion anti-skyrmion pair. In what follows
we shall refer to these field configurations as pseudo-breathers.

In practice, it is very difficult to have a field configuration
of a pseudo-breather. However, we can find approximate field
configurations which still radiate energy and asymptotically reach the
stable (or perhaps only
metastable) configuration configuration of the pseudo-breather.
 The excess of energy over the final 
configuration can then be seen as an excitation energy which is slowly
radiated away. During any scattering process  solitons tend to 
exchange or radiate some energy. When the excitation energy is large enough,
the outgoing pseudo-breather-like configuration
may have  enough energy to evolve into the stable pseudo-breather
field; otherwise,
it ends up with less energy than the metastable configuration and 
progressively dies out. 

The scattering properties of pseudo-breathers are quite interesting.
When the pseudo-breathers are imbedded into the baby-skyrmion model 
the field configurations have an extra degree of freedom 
corresponding to their orientation inside the $\phi_1 , \phi_2$ plane.
When two pseudo-breathers are set at rest near each other,
the force between them depends very much on their relative orientation: when 
they are parallel to each other and oscillate in phase, they
attract each other, overlap and form a new structure which appears to be
an exited pseudo-breather. This pseudo-breather then 
slowly radiates away its energy. The non-topological nature of 
pseudo-breathers means that they can indeed merge to form a new structure
of the same type.

If the two pseudo-breathers are anti parallel, \ie if they oscillate
completely out of phase, then the force between them is repulsive. When the
two pseudo-breathers have a different orientation they slowly rotate
themselves until they become parallel; then they move towards each 
other and form an exited pseudo-breather structure.  

\TwoFigsAB{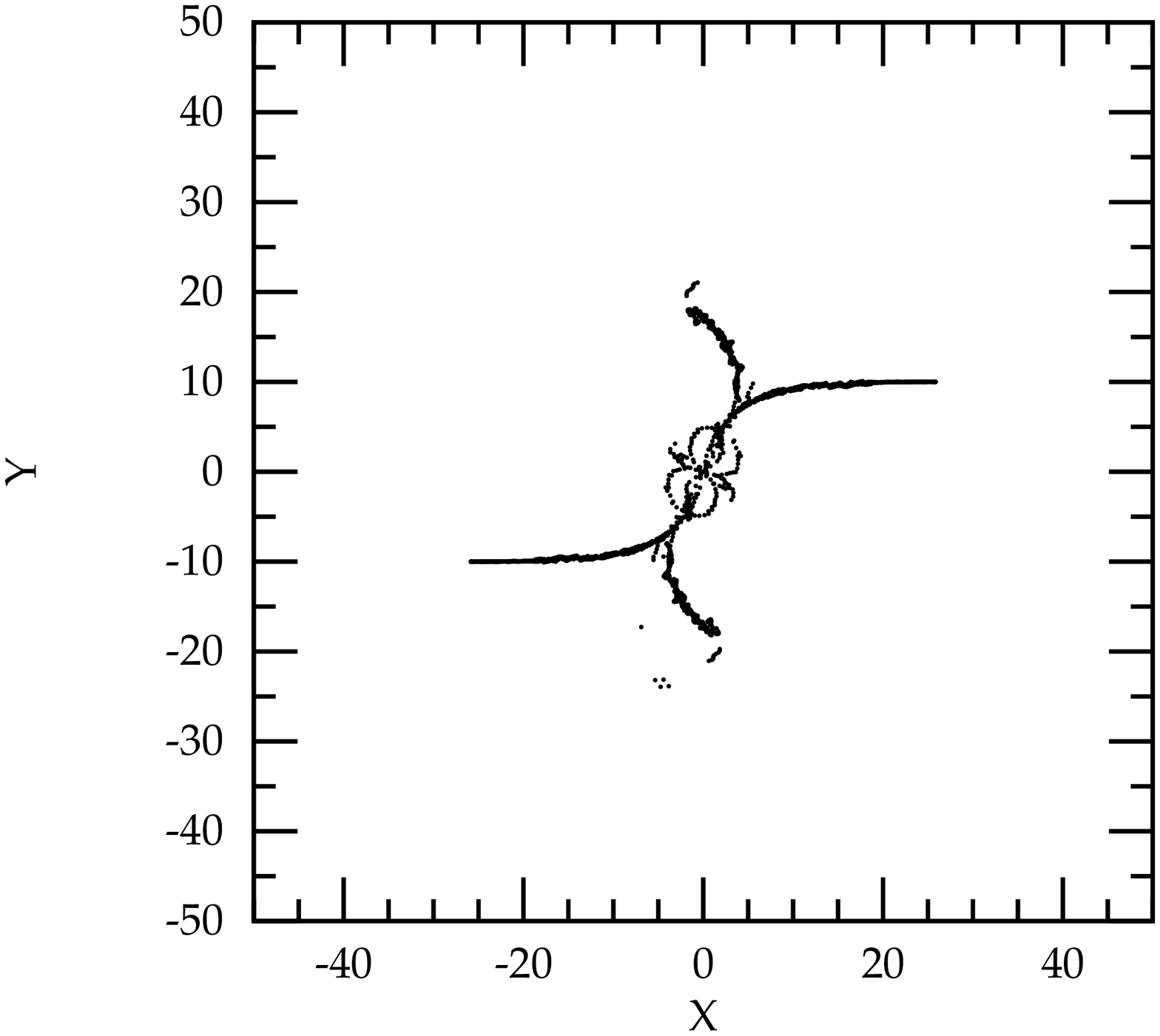}{Pseudo-Breather scattering. Impact Parameter:
10, $v=0.2$} {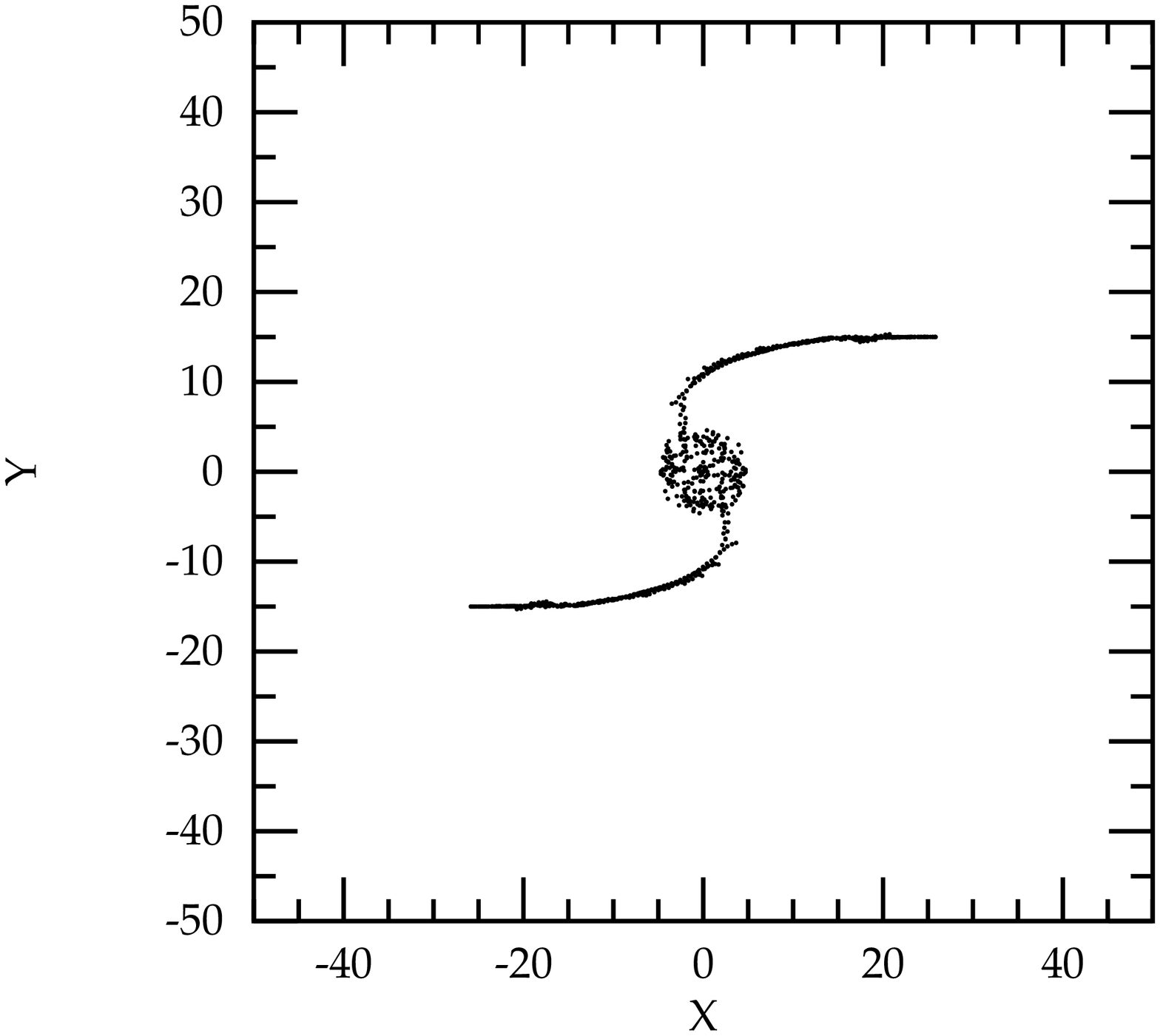}{Pseudo-Breather scattering. Impact Parameter:
15, $v=0.2$}
 
When two  pseudo-breathers are sent towards each other with some kinetic
energy, the scattering is more interesting. Depending on the initial speed or
the scattering impact parameter, they either merge into a single 
pseudo-breather or they undergo a forward scattering. The details 
of these scattering properties are given in [\RTDsinG]. In Figure 7 we show two 
such trajectories. For both scatterings the initial speed was $0.2$, the only 
difference being the values of the impact parameter.


\chapter{Pseudo-Breather-Skyrmion Scattering}
As we have seen, the baby-skyrmion model has two different types of extended
solutions. The skyrmions are topological solitons which are very stable, while 
the pseudo-breathers are time-periodic solitons which can slowly
decay if they are perturbed too much. It is very unusual to have a model that 
exhibits two such different stationary structures and so it is interesting 
to analyse how they interact with each other. 

\OneSizedFig{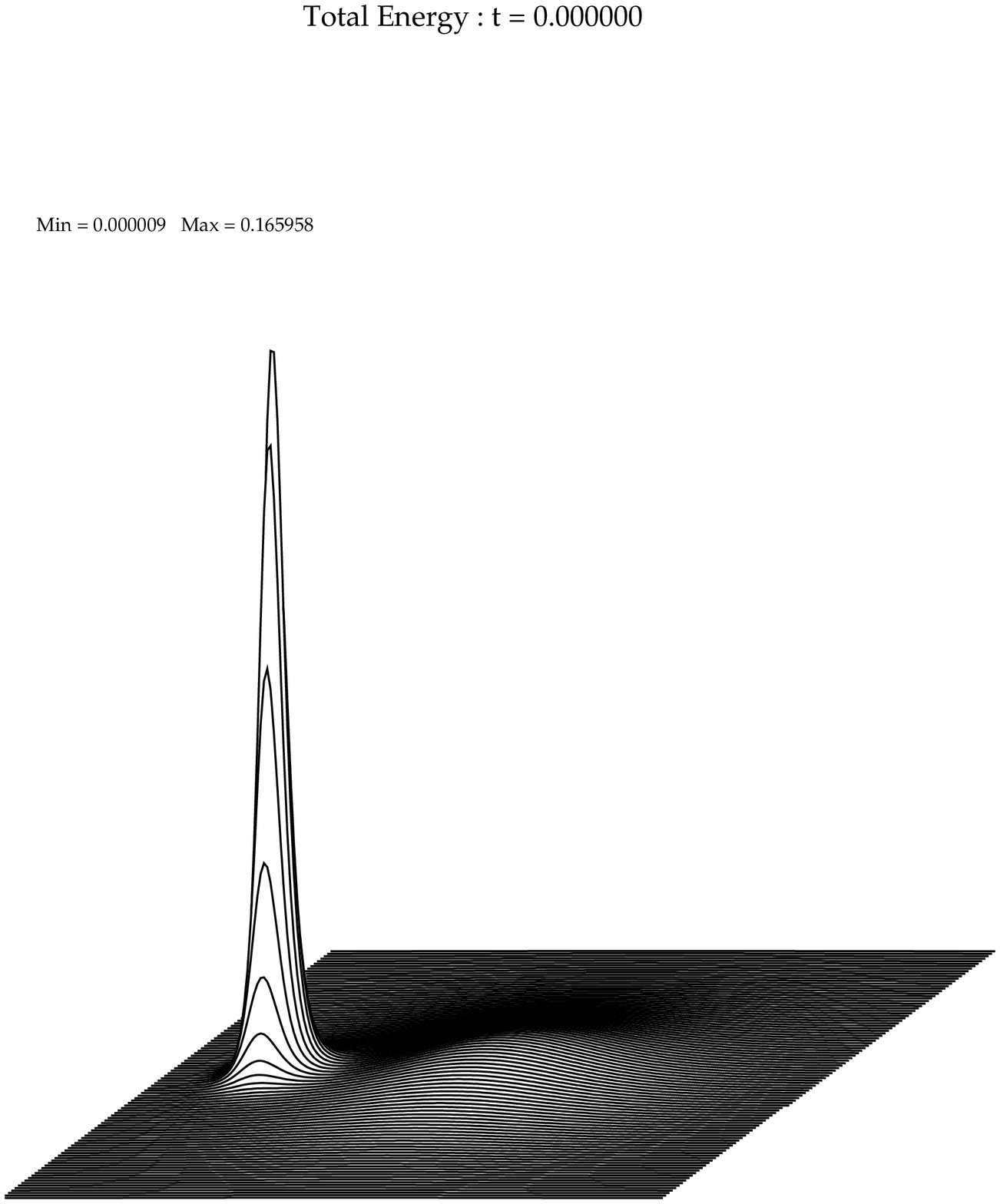}{Skyrmion and pseudo-breather at rest.}{12cm }

When a skyrmion and a pseudo-breather are put at rest next to each other
the overall interaction between them  makes the skyrmion slowly move
away from the pseudo-breather-like configuration 
while the pseudo-breather looses some of its
energy faster than when it is placed there by itself.

To scatter a skyrmion with a pseudo-breather we have placed the 
pseudo-breather soliton at rest, 
and we have sent the skyrmion towards it. We have performed 
this scattering for different orientations of the pseudo-breather, 
for different values of the impact parameter and 
for a few different speeds. In each case, the 
pseudo-breather was initially located at the origin while the skyrmion
always started from $(x_0,y_0)$ where $x_0$ is the initial
position along the $x$ axis and $y_0$ is the impact parameter.
The pseudo-breather was oscillating in the $(\phi_1, \phi_2)$ plane
along the direction $(\Cos(\alpha), \Sin(\alpha))$.

In Figure 8 we show the initial condition corresponding to a baby-skyrmion next
to a pseudo-breather. We note that the skyrmion is much more spiky than the
breather. 


Our numerical results are summarised in the following 4 tables.

\vskip 3mm
\centerline{\vbox{\offinterlineskip\tabskip=0pt\halign
{\strut\vrule#&\quad#\hfill&\vrule#&\hfill\quad#\ &\vrule#&\hfill\quad#\ &\vrule
#&\hfill\quad#\ &\vrule#&\hfill\quad#\ &\vrule#&\hfill\quad#\ 
&\vrule#&\hfill\quad#\ &\vrule#&\hfill\quad#\ &\vrule#\cr
\noalign{\hrule}
& v \quad$\backslash$\quad y0 
     && 15 && 10 && 7.5 && 5 && 3.5 && 2.5 && 1.25 &\cr
\noalign{\hrule}
& 0.2 && 5 && 18 && 31 && 39 && 58 && 67 && -108 &\cr
& 0.3 && 6 && 26 && 16 && 15 && 36 && 35 && -24  &\cr
& 0.4 && 4 &&  7 && 11 &&  0 && -9 &&    && -8   &\cr
\noalign{\hrule}
}}}
\tablecap{Table 1.a : Impact parameter and speed dependance of the scattering angle
($\alpha = 0$, $x_0 = -20$).}

\centerline{\vbox{\offinterlineskip\tabskip=0pt\halign
{\strut\vrule#&\quad#\hfill&\vrule#&\hfill\quad#\ &\vrule#&\hfill\quad#\ &\vrule
#&\hfill\quad#\ &\vrule#&\hfill\quad#\ &\vrule#&\hfill\quad#\ 
&\vrule#&\hfill\quad#\ &\vrule#&\hfill\quad#\ &\vrule#\cr
\noalign{\hrule}
& v \quad$\backslash$\quad y0 
      && 15 && 10 && 7.5 && 5 && 3.5 && 2.5 && 1.25 &\cr
\noalign{\hrule}
& 0.2 && 8 && 19 && 27 && 30 && 21 && 21 && 90 &\cr
& 0.3 && 6 && 16 && 24 && 23 && 21 && 29 && 23 &\cr
& 0.4 && 5 && 20 && 21 && 20 && 18 && 19 && 4  &\cr
\noalign{\hrule}
}}}
\tablecap{Table 1.b : Impact parameter and speed dependance of the scattering angle 
($\alpha = \pi/2$, $x_0 = -20$).}

\centerline{\vbox{\offinterlineskip\tabskip=0pt\halign
{\strut\vrule#&\quad#\hfill&\vrule#&\hfill\quad#\ &\vrule#&\hfill\quad#\ &\vrule
#&\hfill\quad#\ &\vrule#&\hfill\quad#\ &\vrule#&\hfill\quad#\ 
&\vrule#&\hfill\quad#\ &\vrule#&\hfill\quad#\ &\vrule#\cr
\noalign{\hrule}
& v \quad$\backslash$\quad y0
      &&15 && 10 && 7.5&&  5 && 3.5&& 2.5 && 1.25 &\cr
\noalign{\hrule}
& 0.2 && 7 && 18 && 29 && 32 && 52 && 62 && -166 &\cr
& 0.3 && 7 && 14 && 15 && 2.5&& 22 && 45 &&  -27 &\cr
& 0.4 && 3 &&  8 &&  8 && 15 && 14 && 24 &&  -43 &\cr
\noalign{\hrule}
}}}
\tablecap{Table 1.c : Impact parameter and speed dependance of the scattering angle
($\alpha = \pi/4$, $x_0 = -20$).}

\centerline{\vbox{\offinterlineskip\tabskip=0pt\halign
{\strut\vrule#&\quad#\hfill&\vrule#&\hfill\quad#\ &\vrule#&\hfill\quad#\ &\vrule
#&\hfill\quad#\ &\vrule#&\hfill\quad#\ &\vrule#&\hfill\quad#\ 
&\vrule#&\hfill\quad#\ &\vrule#\cr
\noalign{\hrule}
& v \quad$\backslash$\quad x0 
         && 15.2 && 16.2 && 17.2 && 18.2 && 19.2 && 20.2 &\cr
\noalign{\hrule}
& 0.2    && 38   && 33   && 33   &&  38  && 31   && 21   &\cr
& 0.4    && 15   && 16   && 17   &&  20  && 22   && 19   &\cr
\noalign{\hrule}
}}}
\tablecap{Table 2 : Scattering angle as a function of the initial distance 
($\alpha = \pi/2, y_0 = 2.5$).}

The amount of energy lost by the pseudo-breather during the scattering is
larger when  the overlap between the skyrmion and the 
breather, both in time and space, increases. 
In some cases the pseudo-breather is completely destroyed by the collision. 
The oscillation of the pseudo-breather makes the interaction time dependent,
and, as a result, it is difficult to extract a simple pattern from the tables
of scattering angles. 

\chapter{Conclusions}
We have shown that the baby-skyrmion model has many interesting classical 
solutions in addition to  the skyrmion solitons. The
first class of solutions describe exited states of
skyrmions and anti-skyrmions which are unstable with respect to perturbations. 

The second class of solutions involves non-topological stationary 
stationary field configurations which
are periodic in time. They are relatively stable but can be destroyed if they
are perturbed too much.

\ack
One of the authors (AK) thanks The Department for Mathematical Science of 
University of Durham for hospitality during his visit. This visit was 
supported by INTAS grant 93-633 and partly by grants 
INTAS-CNRS1010-CT93-0023 and RFFR-95-02-04681.

We want to thank R.S. Ward for helpful comments.

\refout 
 
\end